\documentstyle[aps,prl,epsfig]{revtex} 
\begin{document}

\title{Comment on ``Evidence for nontrivial ground-state structure of
3d $\pm J$ spin glasses'' }

\author{Anders W. Sandvik}
\address{Department of Physics, University of Illinois 
at Urbana-Champaign, 1110 W.~Green Street, Urbana, Illinois 61801 \\
and Center for Nonlinear Studies, Los Alamos National Laboratory,
Los Alamos, New Mexico 87545}

\date{September 13, 1998}

\maketitle
\vskip3mm


In a recent {\it Letter} \cite{hartmann1}, Hartmann presented results for 
the structure of the degenerate ground states of the three-dimensional $\pm$J 
spin glass model obtained using a genetic algorithm. Further work was carried
out using similar techniques \cite{hartmann2}. In this {\it Comment}, I 
argue that the method does not produce the correct thermodynamic distribution 
of ground states and therefore gives erroneous results for the overlap 
distribution. I present results of simulated annealing \cite{annealing} 
calculations using different annealing rates for cubic lattices with $N=4^3$ 
spins. The disorder-averaged overlap distribution exhibits a significant 
dependence on the annealing rate, even when the energy of the lowest state 
has converged. For fast annealings, moments of the distribution are similar 
to those presented in Ref.~\onlinecite{hartmann1}. However, as the annealing 
rate is lowered, they approach the results obtained by Berg, Hansmann, and 
Celik \cite{berg} using a multi-canonical Monte Carlo method. This shows 
explicitly that care must be taken not only to reach states with the 
lowest energy but also to ensure that they obey the correct thermodynamic
distribution, i.e., that the probability is the same for reaching any of
the ground states.

The simulated annealing procedure was carried out starting at inverse
temperature $\beta=J/T=0$ and increasing $\beta$ in steps of 
$\Delta\beta=0.01$ up to $\beta=3$. For each $\beta$, one or several
Monte Carlo updating cycles were carried out. Each cycle consisted of $N$ 
single-spin Metropolis updates with the spins chosen at random. The 
number of updating cycles per $\beta$ value was chosen as $2^n$, with $n$ 
hence defining the logarithm of the inverse annealing rate. The state with 
the lowest energy last reached in the process was stored. For sufficiently slow
annealings (sufficiently large $n$), this procedure is guaranteed to
give a ground state, with the same probability for reaching all the ground
states. On the order of $3-10\times 10^4$ random configurations with equal 
amounts of $+J$ and $-J$ interactions were studied for $n=0,1,\ldots,5$,
and for each configuration 100 replicas were considered. Out of these,
only those with the lowest energy were used in calculating averages. On 
average, $80\%$ of the replicas reached the lowest energy for 
$n=0$, and for $n=4$ this fraction exceeded $99\%$.

Figure 1 shows the dependence of some disorder-averaged quantities on $n$. 
For the small system size used, the correct ground state energy is 
reproduced already for $n=0$ (i.e., the fastest annealing considered), as 
can be seen in the comparison with the result obtained by P\'al \cite{pal} 
(which also agrees with the less accurate result of Ref.~\cite{hartmann1}). 
However, the moments of the overlap distribution still exhibit a strong 
dependence on $n$. For $n=0$, both the average $\langle |q|\rangle$ and the 
variance $\sigma^2 (|q|)$ of the absolute value of the overlap $q$ are close 
to the results presented in Ref.~\onlinecite{hartmann1}. As $n$ is increased, 
the values change in a way reflecting a reduction in the overlap distribution 
weight for small $q$. They approach the results of multi-canonical Monte 
Carlo simulations \cite{berg}. This behavior indicates that faster annealings 
reach states with the lowest energy but not with the same probability for 
all states. Clearly, this is due to the fast process not exploring the full 
configuration space, and therefore not sufficiently often reaching regions 
where the density of ground states is high. The genetic algorithm
used in Refs.~\onlinecite{hartmann1,hartmann2} can also be expected to be 
affected by such behavior, as there is nothing in the process that guarantees
an equal probability for reaching all ground states. Hence, the results 
obtained with this method do not reflect the thermodynamic behavior of the 
model and the conclusions reached in Ref.~\onlinecite{hartmann1} for 
the ground state structure must be questioned.

I would like to thank N. Hatano and D. Campbell for many useful discussions
and comments on the manuscript. This work was supported by the NSF under 
grant DMR-97-12765.

\begin{figure}
\centerline{\epsfig{width=6in,file=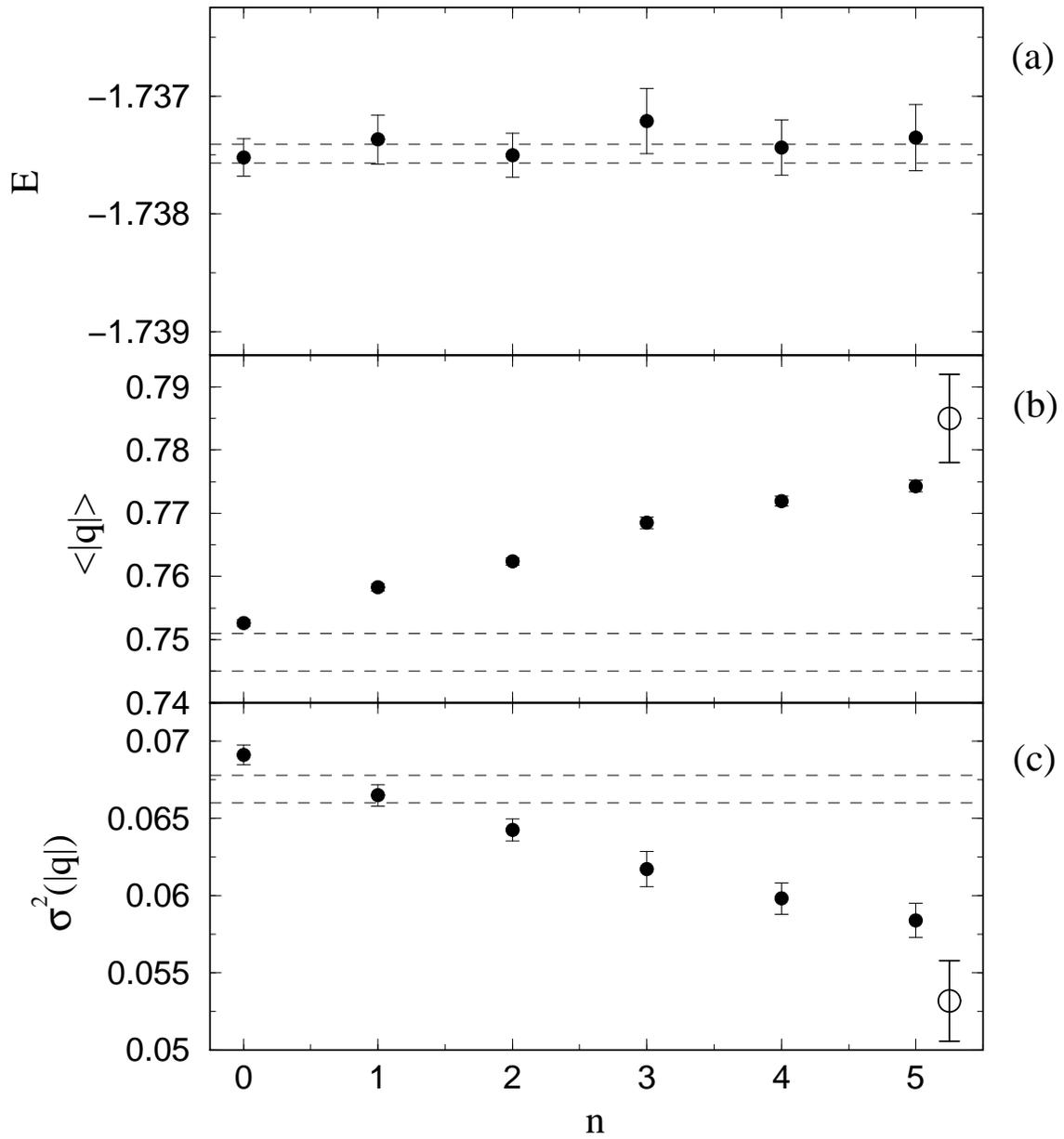}}
\vskip1mm
\caption{Disorder-averaged quantities calculated using lowest-energy
states obtained in simulated annealings with $2^n$ updating cycles per 
temperature. (a) Energy. The result (average $\pm$ one error bar) of Ref.~[5] 
is indicated by the dashed lines. (b) Average and (c) variance of the 
overlap distribution. The results of Ref.~[1] are indicated by the dashed 
lines. The multi-canonical Monte Carlo results [4] are shown as the open
circles.}
\label{fig1}
\end{figure}

\end{document}